\documentclass[onecolumn,sort&compress,numbers]{els-mrw} 

\usepackage{amsmath,amssymb,amsfonts,amsthm,makeidx,graphicx}
\usepackage{txfonts}
\usepackage{helvet}
\usepackage{slashed}

\begin{document}


\chapter{CP violation}\label{chap1}

\author[1]{Alexey A Petrov}%

\address[1]{\orgname{University of South Carolina}, \orgdiv{Department of Physics and Astronomy}, \orgaddress{Columbia, SC 29208, USA}}


\maketitle

\begin{abstract}[Abstract]
CP violation, which involves breaking the combined charge-conjugation (C) and parity (P) symmetries, is essential for understanding the observed matter-antimatter asymmetry in the universe. It is a key feature of the Standard Model (SM), originating from complex phases in the Cabibbo-Kobayashi-Maskawa quark mixing matrix. Despite the successes of the SM, the amount of CP violation it predicts is not enough to explain baryogenesis, prompting searches for new sources of CP violation in other areas of particle physics. This article offers a pedagogical introduction to the theoretical foundations of C, P, and T symmetries and their combinations. It also emphasizes the importance of CP violation in modern particle physics and how it can be used to explore New Physics.
\end{abstract}

\begin{keywords}
 	discrete symmetries\sep CP-violation\sep CPT \sep matter-antimatter asymmetry \sep weak interactions
\end{keywords}

\begin{glossary}[Nomenclature]
	\begin{tabular}{@{}lp{34pc}@{}}
		QFT & Quantum Field Theory\\
		QED & Quantum ElectroDynamics\\
		QCD & Quantum ChromoDynamics\\
		CKM matrix & Cabbibo-Kobayashi-Maskawa matrix\\
		PDG & Particle Data Group\\
		TPC & Triple-Product Correlations
	\end{tabular}
\end{glossary}

\section*{Objectives}
\begin{itemize}
	\item Introduce a concept of charge, parity, and time-reversal transformations in classical and quantum physics
	\item Describe theoretical approaches to C, P, and T symmetries and their roles for the weak, electromagnetic, and strong interactions.
	\item Introduce experimental consequences of conservation and breaking of C, P, and T and combined (CP) symmetries
\end{itemize}


\section{Introduction}\label{intro}

Symmetries are essential for understanding the fundamental interactions of nature. Among them, the discrete symmetries of charge conjugation, parity, and time reversal are especially important in explaining the matter-antimatter imbalance in the universe. These symmetries determine how particles and their interactions behave under specific transformations: (1) Charge conjugation (C), which exchanges a particle with its antiparticle, reversing its charge and other additive quantum numbers; (2) Parity (P), which reflects spatial coordinates; and (3) Time reversal (T), which reverses the direction of time, making a physical process appear as if it were running backward. The way these symmetries are broken, along with their combined actions, such as CP, guides the search for signs of new physics in low-energy experiments.

Studying these symmetries is vital for understanding the observed matter-antimatter asymmetry of the universe \cite{Petrov:2021idw}. According to the Sakharov conditions \cite{Sakharov:1967dj}, any physical theory describing the processes in the universe must meet three fundamental criteria to produce a baryon asymmetry from an initially symmetric universe.
\begin{itemize}
\item
Baryon number violation: interactions that do not conserve baryon number must exist. This is necessary to create a net excess of baryons over antibaryons.
\item
C and CP violation: since charge conjugation symmetry would guarantee equal numbers of baryons and antibaryons, its violation is necessary. Furthermore, combined CP violation is also required to ensure that matter and antimatter behave differently.
\item
Departure from thermal equilibrium: if the universe were in perfect thermal equilibrium, any baryon asymmetry generated by the previous two conditions would be washed out. Nonequilibrium processes, such as those occurring during cosmological phase transitions, are needed to preserve the asymmetry.
\end{itemize}
The known sources of CP violation in the Standard Model, primarily arising from the CKM matrix in the weak interaction, are insufficient to account for the observed asymmetry, requiring the presence of new interactions. 

\section{Classical limit: non-relativistic mechanics}
\label{NonRel}

It would be helpful to introduce C, P, and T transformations in classical systems before analyzing their implications for quantum systems and measurements. If a law of physics -- or a specific interaction or observable -- remains unchanged under these discrete transformations (or any combination of them), it is said that the interaction is {\it invariant} under such a transformation. Invariant interactions can be helpful in constructing models, while the patterns of their breaking can be studied through experimental tests of these models. 

\begin{figure}[h]
	\centering
	\includegraphics[width=8cm,height=5cm]{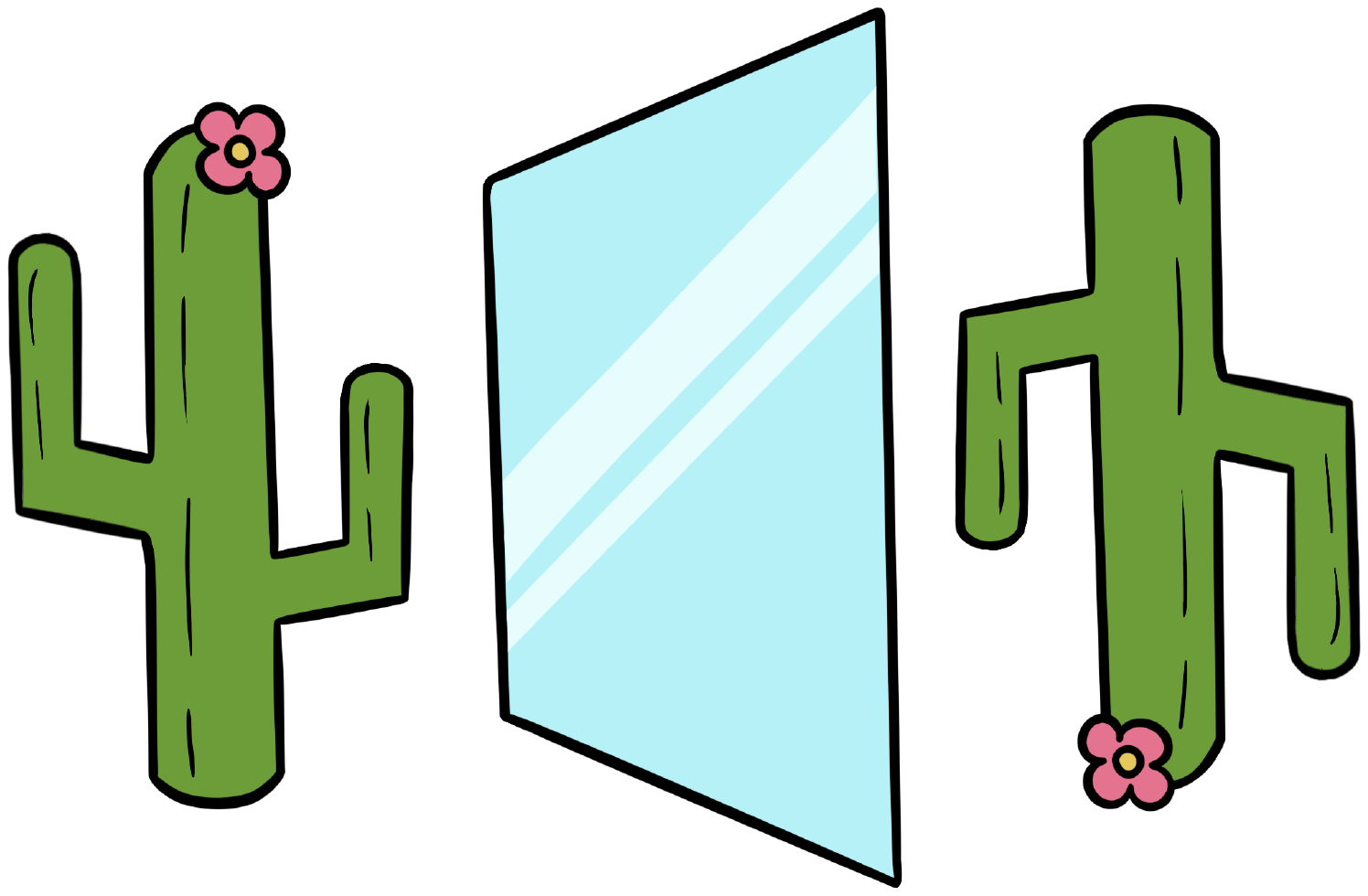}
	\caption{Parity transformation ${\mathbf r} \to -{\mathbf r}$ is equivalent to a reflection through a 
mirror followed by a rotation of $\pi$ radians around an axis defined by the mirror plane (artwork courtesy of Anna A. Petrov)}
	\label{fig:parity}
\end{figure}

Parity, or P-transformation, is a transformation ${\mathbf r} \to -{\mathbf r}$. In the Cartesian basis, it is achieved by inversion of all space coordinates,
\begin{equation}
P: (x, y, z) \to (-x, -y, -z).
\end{equation}
In three dimensions, this transformation is equivalent to a mirror reflection followed by a rotation of $\pi$ radians around an axis defined by the mirror plane, as shown in Fig.~\ref{fig:parity}. This is why this transformation is often called ``reflection,'' even though it does not exactly match how the world looks through a mirror. 

%
Time reversal, or T-transformation, flips the arrow of time, $t \to -t$, 
\begin{equation}
T: t \to -t.
\end{equation}

Charge conjugation, or C-transformation, changes particles to antiparticles, i.e., flips the signs of a particle's charges $q$, including its electric charge,
\begin{equation}
C: q \to -q.
\end{equation}
Since creating a particle-antiparticle pair is a purely quantum process, technically speaking, C-transformation does not belong in classical physics. However, it remains important to introduce this transformation classically, as it will aid in understanding the discussion of discrete symmetries for electromagnetic fields.  

One can analyze how classical quantities change under C, P, and T transformations by using the definitions of various dynamical variables and equations of motion. For example, velocity
\begin{equation}
{\bf v} = \frac{d {\bf r}}{d t},
\end{equation}
is odd under both $P$ and $T$ transformations because $P$ flips the sign of the numerator (${\mathbf r} \to - {\mathbf r}$), while leaving the denominator unchanged. Conversely, the $T$ transformation reverses the sign of the denominator but keeps the numerator intact. This shows that the non-relativistic momentum, ${\mathbf p} = m {\mathbf v}$, is also odd under both $P$ and $T$ transformations. 

This technique allows one to follow ``Ariadne's thread" and examine how other quantities respond to the applications of discrete transformations. For example, one can use Newton's second law to see that the force,
\begin{equation}\label{force}
{\mathbf F} = \frac{d {\mathbf p}}{d t},
\end{equation}
behaves differently under P and T: it is odd under P and even under T, as both numerator and denominator change sign under T. Using the definition of the angular momentum, 
\begin{equation}\label{AngMom}
{\mathbf L} = {\mathbf r} \times {\mathbf p},  
\end{equation}
one can see that it is even under P and odd under T. While the particle's spin represents a purely quantum effect, one should expect that it behaves the same way as angular momentum under those transformations.

Starting from the definition of Eq.~(\ref{force}) regarding how any force behaves under P and T transformation, one can analyze how specific interactions respond to these discrete transformations. In electrodynamics, this can be examined through the definition of the Lorentz force ${\mathbf F}_L$, which describes the motion of an electrically charged particle in electric ${\mathbf E}$ and magnetic $\mathbf{B}$ fields,  
\begin{equation}\label{LorentzForce}
{\mathbf F}_L = q\left(\mathbf{E} + \mathbf{v} \times \mathbf{B}\right),
\end{equation}
where $q$ represents the particle's electric charge. Examining Eq.~(\ref{LorentzForce}) and using what was learned from Eq.~(\ref{force}), one can conclude that under parity transformation $P$
\begin{equation}\label{PEM}
\mathbf{F}_L \xrightarrow{P} -\mathbf{F}_L: \ \ \mathbf{E} \xrightarrow{P} -\mathbf{E}, \quad \mathbf{B} \xrightarrow{P} \mathbf{B} 
\end{equation}
as both ${\mathbf F}_L$ and ${\mathbf v}$ are odd under reflection. At the same time, 
\begin{equation}\label{TEM}
\mathbf{F}_L \xrightarrow{T} \mathbf{F}_L:  \ \ \mathbf{E} \xrightarrow{T} \mathbf{E}, \quad \mathbf{B} \xrightarrow{T} -\mathbf{B}. 
\end{equation}
as ${\mathbf F}_L$ is odd, but ${\mathbf v}$ is even under time reversal transformation.

Now that the electric charge has been considered, one can analyze the effects of a C-transformation on particles and fields. In particular, since C-transformation, by definition, changes $q \to -q$, it is expected that a scalar charge density $\rho({\mathbf r},t)$ also changes sign under C: $\rho({\mathbf r}, t) \to - \rho({\mathbf r}, t)$, and similarly, the current ${\mathbf  j}$ transforms as: ${\mathbf j}({\mathbf r}, t) \to - {\mathbf j}({\mathbf r}, t)$. It should be noted that the continuity equation for the conserved current implies that ${\mathbf j}$ is also odd under P and T transformations. Lastly, one could argue that since C changes the signs of charges—sources of the fields—the fields should also change their signs under C, 
\begin{equation}
 \mathbf{E} \xrightarrow{C} -\mathbf{E}, \quad \mathbf{B} \xrightarrow{C} -\mathbf{B}. 
\end{equation}
\begin{table}
	\TBL{\caption{Maxwell equations and discrete symmetries. The signs show how both sides of a Maxwell equation change under applications of the discrete symmetries and their combinations.}\label{table1}}
	{\begin{tabular*}{\textwidth}{@{\extracolsep{\fill}}@{}lcccc@{}}
			\toprule
			\multicolumn{1}{@{}l}{\TCH{Equation}} &
			\multicolumn{1}{c}{\TCH{$P$}} &
			\multicolumn{1}{c}{\TCH{$T$}}  &
			\multicolumn{1}{c}{\TCH{$C$}} &
			\multicolumn{1}{c}{\TCH{$CPT$}}\\
			\colrule
			$\boldmath \nabla \cdot E = 4\pi \rho$ & $+$ & $+$ & $-$ & $-$ \\
			$\nabla \cdot B = 0$ & $-$ & $-$ & $-$ & $-$\\
			$\nabla \times B - \frac{\partial E}{\partial t} = 4\pi j$ & $-$ & $-$ & $-$ & $-$\\
			$\nabla \times E + \frac{\partial B}{\partial t} =0$ & $+$ & $+$ & $-$ & $-$\\
			\botrule
	\end{tabular*}}{
}
\end{table}
Armed with what has been learned so far, one can now see how the laws of electrodynamics, i.e., the Maxwell equations, behave under C, P, T, and their combinations. The results are available in Table~\ref{table1}. As can be seen, in each case, both sides of Maxwell's equations exhibit the same transformation properties under all discrete symmetries. Therefore, no $CP$ violation can be introduced at the classical level.

It is also helpful to remember that mechanical and quantum-mechanical descriptions of electromagnetic interactions of charged particles are most conveniently expressed in terms of potentials, which are conventionally introduced as
\begin{eqnarray}
{\bf E} &=& - {\bf \nabla} \varphi - \frac{\partial {\mathbf A}}{\partial t}, 
\nonumber \\
{\bf B} &=& {\bf \nabla} \times {\mathbf A},
\end{eqnarray}
where $\varphi$ is a scalar potential, and ${\mathbf A}$ is a vector potential. These definitions determine how the potentials respond to the action of discrete symmetry transformations. For example, under parity
\begin{eqnarray}
  \varphi  \xrightarrow{P} \varphi, \quad \  \mathbf{A}   \xrightarrow{P} - \mathbf{A},
\end{eqnarray}
as ${\mathbf \nabla} \equiv \partial/\partial {\mathbf r} \xrightarrow{P} - {\mathbf \nabla}$ and following Eq.~(\ref{PEM}). Similarly, using Eq.~(\ref{TEM}) and the fact that $\partial/\partial t \to - \partial/\partial t$, one can see that 
\begin{eqnarray}
  \varphi  \xrightarrow{T} \varphi, \quad \  \mathbf{A}   \xrightarrow{T} - \mathbf{A}.
\end{eqnarray}
It is interesting to observe that different vectorial quantities transform differently under parity: as can be seen from Eq.~(\ref{PEM}), ${\mathbf E}$ transforms just like ${\mathbf r}$, while ${\mathbf B}$ does not. Because of that, vectorial quantities that transform like ${\mathbf r}$ under parity are called {\it polar} vectors (or simply vectors), while the others (like ${\mathbf B}$) are called {\it axial} or {\it pseudo} vectors. Similar statements can be made regarding the scalar type of quantities: while ${\mathbf r}^2 = {\mathbf r}\cdot {\mathbf r}$ stays invariant under $P$, the $({\mathbf E}\cdot {\mathbf B})$ flips the sign under P. It is because of this behavior that the later quantities are called {\it pseudoscalars}. An example of such a quantity is helicity.

The concept of helicity $h$, a projection of a particle's spin along the direction of its momentum, is used extensively in particle physics. It is defined as 
\begin{equation}
h = \frac{{\mathbf s} \cdot {\mathbf p}}{|{\mathbf s}||{\mathbf p}|},
\end{equation}
From its definition, one can see that helicity is a frame-dependent concept. For example, it flips its sign if one chooses a frame in which the observer moves faster than the particle. Yet, it is a good quantity to define for the massless particles as they move with the speed of light. In this case, the helicity is a conserved quantum number and is equivalent to chirality. While helicity is related to spin, i.e., it is not a classical quantity, we can still infer its transformation properties from what we know about spin and momentum in classical physics. It can be seen that $h \to h$ under C, $h \to h$ under T, and $h \to -h$ under P, which makes it a pseudoscalar.

\section{Classical limit: relativistic mechanics and electrodynamics}
\label{Rel}

To introduce C, P, and T transformations in relativistic quantum field theory, it is essential to understand how relativistic objects, such as four-vectors, transform under discrete symmetry operations. A coordinate four-vector is defined as $x^\mu = (t, x, y, z)$, where $\mu = 0,1,2,3$, and the speed of light $c=1$. Using the definitions from Section \ref{NonRel}, the mathematical expression for the effect of a parity transformation can be written as
\begin{equation}
P: (t, x, y, z) \to (t, -x, -y, -z), 
\end{equation}
which implies for the position four-vector $x^\mu$
\begin{equation}
 x^\mu = (t, \mathbf{r})  \xrightarrow{P}  x_\mu = (t, -\mathbf{r}),
\end{equation}
i.e., the parity transformation changes a contravariant vector into a covariant one and vice versa. Since the velocity $u^\mu$ is the derivative of position, it transforms under P similarly, $u^\mu \to u_\mu$. The energy-momentum four-vector also transforms this way: $p^\mu = (E, \mathbf{p})$ changes to $(E, -\mathbf{p})$, meaning energy remains unchanged while momentum changes its sign.

The electromagnetic field tensor $F^{\mu\nu}$ transforms according to the way the electric field $\mathbf{E}$ and magnetic field $\mathbf{B}$ transform. From Eq.~(\ref{PEM}),
\begin{equation}\label{FieldTensor}
F^{\mu\nu} = \begin{bmatrix}
0 & -E_x & -E_y & -E_z \\
E_x & 0 & -B_z & B_y \\
E_y & B_z & 0 & B_x \\
E_z & - B_y & B_x & 0
\end{bmatrix} \quad
\xrightarrow{P}  \quad 
\begin{bmatrix}
0 & E_x & E_y & E_z \\
-E_x & 0 & -B_z & B_y \\
-E_y & B_z & 0 & B_x \\
-E_z & - B_y & B_x & 0
\end{bmatrix}
= F_{\mu\nu} .
\end{equation}
Or, in the four-dimensional formulation, the field tensor components transform under parity as
\begin{equation}
F^{0i} \to -F^{0i}, \quad F^{ij} \to F^{ij},
\end{equation}
where the indices $i, j$ refer to spatial components. This ensures that Gauss’s and Ampere’s laws retain their forms, provided the current density transforms as $j^i \to -j^i$.

The time reversal involves reversing the direction of time, so, using the definitions from Section \ref{NonRel}, the effect of this transformation can be expressed as
\begin{equation}
T: (t, x, y, z) \to (-t, x, y, z).
\end{equation}
which implies that 
\begin{equation}
 x^\mu = (t, \mathbf{r})  \xrightarrow{T}  -x_\mu = (-t, \mathbf{r}),
\end{equation}
The velocity four-vector changes sign due to the derivative with respect to (proper) time. The energy-momentum four-vector transforms as $p^\mu = (E, \mathbf{p}) \to (E, -\mathbf{p})$, with 3-momentum reversing and energy remaining unchanged (depending on the context). Using the definition on the left-hand of the Eq.~(\ref{FieldTensor}), one can infer that the components of $F^{\mu\nu}$ transform under time-reversal as
\begin{equation}
F^{0i} \to F^{0i}, \quad F^{ij} \to -F^{ij}.
\end{equation}
This transformation maintains the invariance of Maxwell’s equations, as long as the current density transforms with $j^0 \to j^0$ and $j^i \to -j^i$, thereby preserving charge conservation and the structure of electromagnetic wave equations. Specifically, the relativistic equations of motion in mechanics and electrodynamics are generally invariant under these discrete transformations. For instance, the relativistic formulation of Newton’s second law, 
\begin{equation}
F^\mu = \frac{d p^\mu}{d \tau}, 
\end{equation}
where $\tau$ is the proper time, remains invariant under these transformations, as the transformation properties of force, four-momentum, and acceleration maintain consistency with the governing equations. For a charged particle moving in an electromagnetic field, the Lorentz force equation is given by:
\begin{equation}
\frac{d p^\mu}{d \tau} = q F^{\mu\nu} u_\nu.
 \end{equation}
Under parity, the transformation of the four-momentum $p^\mu$ and the electromagnetic field tensor $F^{\mu\nu}$ ensures that both sides of the equation transform identically, preserving its form. Here, the charge $q$ is a scalar that remains unchanged under parity.

For time-reversal, the four-velocity $u^\mu$ transforms as $u^0 \to u^0, \mathbf{u} \to -\mathbf{u}$, leading to the transformation $p^\mu \to (E, -\mathbf{p})$. The electromagnetic field components transform as $\mathbf{E} \to \mathbf{E}$ and $\mathbf{B} \to -\mathbf{B}$, modifying the Lorentz force equation. The sign reversal of $\mathbf{u}$ and $\mathbf{B}$ ensures that the cross-product term  $\mathbf{u} \times \mathbf{B}$ remains consistent under time-reversal, preserving the equation’s structure.

Maxwell’s equations exhibit invariance under these discrete transformations when charge and current densities transform appropriately. The continuity equation, $\partial_\mu j^\mu = 0$, remains unchanged under both P and T transformations, ensuring charge conservation. It might be helpful to remember that $\partial^\mu \to \partial_\mu$ under parity, and $\partial^\mu \to -\partial_\mu$ under time reversal.

The transformation properties of four-vectors and field tensors under C, P, and T symmetries guarantee the consistency and invariance of fundamental physical laws. While classical theories respect these symmetries, Sakharov conditions require breaking some of them in quantum mechanics. 

\section{C, P, and T in quantum mechanics: nonrelativistic case}

In quantum mechanics, it is necessary to modify the concepts of C, P, and T transformations. This is because symmetry transformations in quantum mechanics are implemented through the actions of unitary operators. Therefore, C and P are promoted to unitary operators ${\cal C}$ and ${\cal P}$. 

The parity operator ${\cal P}$  acts on the wavefunction $\psi(\mathbf{r})$ by changing the coordinates ${\mathbf r} \to - {\mathbf r}$,
\begin{equation}
{\cal P} \ \psi(\mathbf{r}) = \psi(-\mathbf{r}).
\end{equation}
Applying the parity transformation again, one returns to the original state described by the same $\psi(\mathbf{r})$. This shows that ${\cal P}^2 = I$, implying that ${\cal P}$ is a unitary operator.

Similarly to classical physics, charge conjugation is not naturally defined in nonrelativistic quantum mechanics since antiparticles do not emerge at nonrelativistic energies. Yet, one can still introduce charge transformations in simplified systems. For example, one can consider a two-level system, with one state describing the ``particle" ($|+\rangle$) and the second state describing the ``antiparticle" ($|-\rangle$). In such a system, one can define a charge conjugation operator ${\cal C}$ such that
\begin{equation}
 {\cal C} \ |+\rangle = |-\rangle, \quad {\cal C} \ |-\rangle = |+\rangle
\end{equation}
Applying the charge conjugation operator twice, one returns to the original state. This shows that ${\cal C}^2 = I$, implying that ${\cal C}$ is a unitary operator. Also,
\begin{equation}
{\cal C}^\dagger = {\cal C}^{-1}, \ {\cal P}^\dagger = {\cal P}^{-1}.
\end{equation}
If those transformations preserve the laws of physics and are good symmetries of Nature, the Sommerfield theorem \cite{Sakurai:2011zz} requires that the operators that represent such transformations commute with the Hamiltonian ${\cal H}$ of the quantum-mechanical system,
\begin{equation}
\left[ {\cal C}, {\cal H}\right] = 0, \ \left[ {\cal P}, {\cal H}\right] =0.
\end{equation}
This implies that, for example, the solutions of the Schr\"odinger equation, or the energy eigenstates, can be chosen to have definite parity: they are either even (symmetric) or odd (antisymmetric) under spatial inversion.

This could be a good point to proceed by constructing the explicit form of those operators. However, one must pause to realize that the statement above is not entirely accurate. It is known experimentally \cite{Wu:1957my} that weak interactions do not respect both $C$ and $P$! In this case,
\begin{equation}
\left[ {\cal C}, {\cal H}_{\rm W}\right] \neq 0, \ \left[ {\cal P}, {\cal H}_{\rm W}\right] \neq 0,
\end{equation}
So, parity and charge conjugation cannot be considered good quantum numbers! To address this problem, one must rely on the fact that weak interactions are short-range (and weak) to prepare the initial and final states in states with definite P and C. This allows the construction of (interpolating) operators with well-defined P and C quantum numbers.

The above discussion concerns the properties of the unitary operators corresponding to ${\cal P}$ and ${\cal C}$ transformations. How does one define the time-reversal operator ${\cal T}$? The discussion similar to that of the ${\cal P}$ operator does not apply directly, as can be clearly seen by examining the Schr"odinger equation,
\begin{equation}\label{schro}
i\frac{\partial \psi}{\partial t} = - \frac{\vec \nabla^2}{2 m} \psi,
\end{equation}
where, if one applies the concepts introduced in Sections \ref{NonRel} and \ref{Rel}, one realizes that the left-hand side of Eq.~(\ref{schro}) is odd, while the right-hand side is even! 

Since one should be able to define the time reversal operation in quantum mechanics, a helpful observation is that if one requires that time reversal also involves changing $i \to -i$ and $\psi \to \psi^*$, the Schr\"odinger equation retains its form. This leads to the conclusion that, in contrast to P and C, the T-transformation is achieved through an {\it anti-unitary} transformation\footnote{Recall that antiunitary transformation ${\hat A}$ can be thought of as a combination of a unitary transformation ${\hat U}$ and complex conjugation ${\hat K}$, i.e., ${\hat A}={\hat U} {\hat K}$ \cite{Sakurai:2011zz}.}. The time-reversal operator possesses several valuable properties. For instance, the classical time reversal transformation maintains the constancy of energy. This indicates that the time reversal operator ${\cal T}$ commutes with the Hamiltonian,
\begin{equation}\label{TR_Ham}
\left[{\cal T}, {\cal H}\right] = 0 \ \ \mbox{or} \ \ {\cal T} {\cal H} {\cal T} ^{-1} = {\cal H}.
\end{equation}
This results in a different action of a time reversal on a scattering matrix $S$ \footnote{This follows from the definition of the scattering matrix as $S \equiv \lim_{t\to \infty}\lim_{t'\to \infty} \exp{(-i{\cal H}(t-t'))}$},
\begin{equation}
{\cal C} S {\cal C}^{-1} = S, \ \  {\cal P} S {\cal P}^{-1} = S, \ \ 
{\cal T} S {\cal T}^{-1} = S^\dagger.
\end{equation}
In other words, a time reversal transformation also swaps the in- and out-states of the S-matrix.

\section{C, P, and T in relativistic quantum mechanics and quantum field theory}

In relativistic quantum mechanics, the concept of charge conjugation can be appropriately introduced since relativistic interactions can produce particle-antiparticle pairs. This is where charge conjugation begins to play an important role. If particles and antiparticles have the same mass and opposite charge\footnote{Originally, this assumption originated from Dirac's hole theory of a positron, where positrons were envisioned as ``holes'' in the sea of filled electrons states.}, one must ensure that the Dirac equation accommodates a new symmetry that describes the exchange of particles and antiparticles. In other words, if there is a wave function describing a state $\psi$,
\begin{equation}\label{DiracEquation}
\left(i\slashed \partial - e \slashed A - m\right) \psi = 0,
\end{equation}
there must be a wave function of the charge-conjugated state $\psi^c$ that satisfies the same Dirac equation but with the opposite charge,
\begin{equation}\label{DiracEquationConj}
\left(i\slashed \partial + e \slashed A - m\right) \psi^c = 0.
\end{equation}
The equation for $\psi^c$ can be constructed by manipulating Eq.~(\ref{DiracEquation}). By taking the complex conjugate and transposing, 
\begin{equation}\label{DiracEquationConj2}
\left[\gamma^{\mu T} \left(-i\partial_\mu - e A_\mu\right) - m\right] \gamma^{0 T} \psi^* = 
\left[\gamma^{\mu T} \left(-i\partial_\mu - e A_\mu\right) - m\right] {\overline \psi}^T= 0,
\end{equation}
where the short-hand notation $\gamma^{0 T} \psi^* = {\overline \psi}^T$ is also introduced. Applying ${\cal C}$ on the left in Eq.~(\ref{DiracEquationConj2}) and utilizing the relationship ${\cal C} {\cal C}^{-1}={\cal C}^{-1}{\cal C}=1$, 
\begin{equation}\label{DiracEquationConj3}
{\cal C} \left[\gamma^{\mu T} {\cal C}^{-1} {\cal C} \left(-i\partial_\mu - e A_\mu\right) - m\right] {\cal C}^{-1} {\cal C} {\overline \psi}^T= 
\left[{\cal C} \gamma^{\mu T} {\cal C}^{-1} \left(-i\partial_\mu - e A_\mu\right) - m\right] {\cal C} {\overline \psi}^T = 0.
\end{equation}
Compared to Eq.~(\ref{DiracEquationConj}), it is clear that the following must be true for the Dirac equation to properly describe the positron state,
\begin{equation}
{\cal C} \gamma^{\mu T} {\cal C}^{-1} = - \gamma^\mu.
\end{equation}
While the exact form of the matrix for the operator ${\cal C}$ that satisfies this condition depends on the chosen representation of the Dirac matrices, such matrices can always be found. For example, in the Dirac representation, this matrix takes the form ${\cal C} = i\gamma^2 \ gamma^0$. It also follows from Eq.~(\ref{DiracEquationConj3}) that
\begin{equation}
\psi^c = \eta_c {\cal C} {\overline \psi}^T, 
\end{equation}
where an arbitrary phase $\eta_c = e^{i \beta_{\rm C}}$ has been introduced to indicate that the wave function is defined only up to a phase. A convention is often used with $\eta_c=1$. It is noteworthy that the Dirac equation in Eq.~(\ref{DiracEquationConj}) remains invariant under the combined transformation $\psi \to \psi^c =  \eta_c {\cal C} {\overline \psi}^T$ and $A_\mu \to A_\mu^c = - A_\mu$. 

In quantum field theory, fields are also represented by operators, allowing one to see how C, P, and T act on them. To do this, it might help to keep in mind the mode decomposition of the field operators. This is easiest to visualize for scalar fields,  
\begin{equation}
\phi(x) = \int \frac{d^3 p}{(2\pi)^3 2E} \sum_s \left[ a_{\bf p} e^{-i p\cdot x}
+  b_{\bf -p}^\dagger e^{i p\cdot x} \right]
\end{equation}
where $a_{\bf p}$ and $b_{\bf p}^\dagger$ are creation and annihilation operators. The one-particle states can be generated by applying the creation operators to the vacuum state. 
\begin{equation}
|{\bf p} \rangle = \sqrt{2E} a_{\bf p}^\dagger |0\rangle.
\end{equation}
We expect that under parity,
\begin{equation}\label{ScalarP}
{\cal P} \phi(t, \mathbf{r}) {\cal P}^\dagger = \eta^\phi_p \phi(t, -\mathbf{r}), 
\end{equation}
where $\eta^\phi_p = \exp{(i\alpha_p)}$ is an arbitrary phase. For the time-reversal transformation, one needs to flip the sign of the time variable, 
\begin{equation}\label{ScalarT}
{\cal T} \phi(t, \mathbf{r}) {\cal T}^\dagger = \eta^\phi_t \phi(-t, \mathbf{r}), 
\end{equation}
where $\eta^\phi_t = \exp{(i\alpha_t)}$ is, again, an arbitrary phase. Under charge conjugation,
\begin{equation}\label{ScalarC}
{\cal C} \phi(t, \mathbf{r}) {\cal C}^\dagger = \eta^\phi_c \phi^\dagger(t, \mathbf{r}), 
\end{equation}
with $\eta^\phi_c = \exp{(i\alpha_c)}$ being an arbitrary phase. The mode decomposition enables the determination of the effects of parity and other discrete transformations on creation and annihilation operators. 

For the fermion operators, the actions of P, T, and C are more complicated, as they involve analyzing how spinors change under those transformations.
\begin{eqnarray}
\psi(x) &=& \int \frac{d^3 p}{(2\pi)^3 2E} \sum_s \left[ a_{\bf p}^s u_s(p)e^{-i p\cdot x}
+  b_{\bf p}^{s \dagger} v^s(p) e^{i p\cdot x} \right], \nonumber \\
\overline \psi(x) &=& \int \frac{d^3 p}{(2\pi)^3 2E} \sum_s \left[ b_{\bf p}^s \bar u_s(p)e^{-i p\cdot x}
+  a_{\bf p}^{s \dagger} v^s(p) e^{i p\cdot x} \right],
\end{eqnarray}
where $a_{\bf p}^{s} (b_{\bf p}^s)$ and $a_{\bf p}^{s \dagger} (b_{\bf p}^{s \dagger})$ are again creation and annihilation operators, and $u$ and $v$ represent the 4-spinors. The one-particle states can also be obtained by applying the creation operators to the vacuum state. For example, an electron state can be 
\begin{equation}
|e^-({\bf p},s) \rangle = \sqrt{2E} a_{\bf p}^{s \dagger} |0\rangle.
\end{equation}
Note that these definitions follow the normalization of states used in \cite{Peskin:1995ev}.

With these mode decompositions, one can study the transformations of field operators under C, P, and T and their combinations. Derivations of such transformations are integral parts of standard courses in quantum field theory \cite{Peskin:1995ev,Branco:1999fs}. It would be useful to list them here for completeness. The actions of all C, P, and T transformations are derived in \cite{Branco:1999fs}. 

For the scalar fields,
\begin{eqnarray}\label{scalCPT}
{\cal P} \phi(t, {\mathbf r}) {\cal P}^\dagger &=& e^{i \alpha_{\rm P}} \phi(t, -{\mathbf  r}), 
\nonumber \\
{\cal C} \phi(t, {\mathbf  r}) {\cal C}^\dagger &=& e^{i \alpha_{\rm C}} \phi^\dagger(t, {\mathbf  r}),
\\
{\cal T} \phi(t, {\mathbf  r}) {\cal T}^{-1} &=& e^{i \alpha_{\rm T}} \phi(-t, {\mathbf  r}), 
\nonumber
\end{eqnarray}
where $\alpha_i$ are arbitrary phases. For the vector fields, in particular the QED photon $A_\mu$,
\begin{eqnarray}\label{vecCPT}
{\cal P} A_\mu(t, {\mathbf  r}) {\cal P}^\dagger &=& A^\mu(t, -{\mathbf  r}), 
\nonumber \\
{\cal C} A_\mu(t, {\mathbf  r}) {\cal C}^\dagger &=& -A_\mu(t, {\mathbf  r}),
\\
{\cal T} A_\mu(t, {\mathbf  r}) {\cal T}^{-1} &=& A^\mu(-t, {\mathbf  r}),
\nonumber
\end{eqnarray}
as $A_\mu$ is a Hermitian operator. Finally, for the fermion fields,
\begin{eqnarray}\label{diracCPT}
{\cal P} \psi(t, {\mathbf  r}) {\cal P}^\dagger &=& e^{i \beta_{\rm P}} \gamma^0 \psi(t, -{\mathbf  r}), 
\nonumber \\
{\cal C} \psi(t, {\mathbf r}) {\cal C}^\dagger &=& e^{i \beta_{\rm C}} \psi^c(t, {\mathbf r}),
\\
{\cal T} \psi(t, {\mathbf r}) {\cal T}^{-1} &=& e^{i \beta_{\rm T}} \gamma_0^* \gamma_5^* C^* A \psi(-t, {\mathbf r}), 
\nonumber
\end{eqnarray}
where, again, $\beta_i$ are the arbitrary phases, and $\psi^c = C A^T {\psi^\dagger}^T$ and $A=\gamma^0$ for Dirac or Majorana representation of the Direct matrices \cite{Branco:1999fs}.

While the field transformations described by Eqs.~(\ref{scalCPT}), (\ref{vecCPT}), and (\ref{diracCPT}) are helpful, greater interest lies in the Lorentz-covariant objects constructed from scalar, fermion, and vector fields, as well as their responses to C, P,  and T transformations. These objects serve as the main building blocks of SM and BSM Lagrangians and are therefore of practical significance. The results are summarized in Table \ref{table2}.
\begin{table}
	\TBL{\caption{Discrete symmetries and fermionic currents. Here $\psi$ and $\chi$ represent fermion fields. }\label{table2}}
	{\begin{tabular*}{\textwidth}{@{\extracolsep{\fill}}@{}lccccc@{}}
			\toprule
			\multicolumn{1}{@{}l}{\TCH{Current}} &
			\multicolumn{1}{c}{\TCH{$P$}} &
			\multicolumn{1}{c}{\TCH{$T$}}  &
			\multicolumn{1}{c}{\TCH{$C$}} &
			\multicolumn{1}{c}{\TCH{$CP$}} &
			\multicolumn{1}{c}{\TCH{$CPT$}}\\
			\colrule
$\overline \psi \chi$ & $\overline \psi \chi$ & $\overline \psi \chi$ & $\overline \chi \psi$ & $\overline \chi \psi$ & $\overline \chi \psi$
\\
$\overline \psi \gamma_5 \chi$ & $-\overline \psi \gamma_5 \chi$ & $\overline \psi \gamma_5 \chi$ & $\overline \chi \gamma_5 \psi$ & 
$-\overline \chi \gamma_5 \psi$ & $-\overline \chi \gamma_5 \psi$
\\
$\overline \psi \gamma_\mu \chi$ & $\overline \psi \gamma_\mu \chi$ & $\overline \psi \gamma_\mu \chi$ & 
$-\overline \chi \gamma_\mu \psi$ & $-\overline \chi \gamma_\mu \psi$ & $-\overline \chi \gamma_\mu \psi$
\\
$\overline \psi  \gamma_\mu \gamma_5 \chi$ & $-\overline \psi  \gamma_\mu \gamma_5 \chi$ & 
$\overline \psi  \gamma_\mu \gamma_5 \chi$ & $\overline \chi  \gamma_\mu \gamma_5 \psi$ & $-\overline \chi  \gamma_\mu \gamma_5 \psi$ & 
$-\overline \chi  \gamma_\mu \gamma_5 \psi$
\\
$\overline \psi \sigma_{\mu\nu} \chi$ & $\overline \psi \sigma_{\mu\nu} \chi$ & $-\overline \psi \sigma_{\mu\nu} \chi$ & 
$-\overline \chi \sigma_{\mu\nu} \psi$ & $-\overline \chi \sigma_{\mu\nu} \psi$ & $\overline \chi \sigma_{\mu\nu} \psi$
\\
			\botrule
	\end{tabular*}}{%
		\begin{tablenotes}
			\footnotetext{\source{See Ref.~\cite{Branco:1999fs}}}
		\end{tablenotes}
	}%
\end{table}

The information presented in Table \ref{table2} can be used to devise how left- and right-handed currents respond to discrete transformations. For example, for a vector current of left-handed fields, 
\begin{equation}
j_{L, \mu} = \overline \psi_L \gamma_\mu \chi_L = \frac{1}{2} \left(
\overline \psi \gamma_\mu \chi - \overline \psi \gamma_\mu \gamma_5 \chi
\right),
\end{equation}
from which it follows that it transforms into itself under ${\cal T}$ and into a right-handed current $j_{R, \mu} = \overline \psi_R \gamma_\mu \chi_R$ under ${\cal P}$. It also transforms into 
a negative of a Hermitian-conjugated version of $j_{R, \mu}$ under ${\cal C}$. Using Eqs.~(\ref{scalCPT}), (\ref{vecCPT}), and (\ref{diracCPT}), as well as 
Table \ref{table2}, transformation laws of all of the terms in a Lagrangian can be obtained. One needs to remember that under space reflections, one should have $\partial^\mu \to \partial_\mu$, and under time reversal, one should expect that $\partial^\mu \to -\partial_\mu$.

\section{Applications}
\label{Applications}

\subsection{Kramer's degeneracy}
\label{Kramers}

There is an interesting consequence of the time-reversal invariance for a system of fermions in quantum mechanics. While ${\cal T}$, being antiunitary, cannot correspond to a conserved quantity,  ${\cal T}^2$, in fact, can be associated with a quantum number,
\begin{equation}
{\cal T}^2 = {\cal T} {\cal T} = \hat U \hat K \hat U \hat K =   \hat U  \hat U^*,
\end{equation}
as it is unitary! Here, a representation of an anti-unitary operator ${\cal }T \equiv \hat U \hat K$, where $\hat K$  is a complex conjugation operator, has been employed. It can be demonstrated \cite{Branco:1999fs} that ${\cal T}^2$ has two possible eigenvalues: $+1$ for the bosonic systems and $-1$ for the fermionic systems. This observation leads to an intriguing theorem commonly referred to as {\it Kramer's degeneracy}.

\begin{theorem}[Kramer's degeneracy]
\textit{In a time-reversal invariant fermionic system, all energy eigenstates are (at least) doubly degenerate.}
\end{theorem}

\begin{proof}
Let $|E\rangle$ be an eigenstate of a Hamiltonian of the system of fermions ${\cal H}$,
\begin{equation}
{\cal H}|E\rangle = E|E\rangle.
\end{equation}
Acting on this equation with the operator ${\cal T}$ from the left, one can see that 
\begin{eqnarray}
{\cal T H}|E\rangle &=& {\cal T} E |E\rangle = E ({\cal T}|E\rangle) 
\nonumber \\
&=& {\cal H T}|E\rangle =  E|E_T\rangle,
\end{eqnarray}
implying that $|E_T\rangle \equiv {\cal T}| E\rangle$ is also an eigenstate of the Hamiltonian ${\cal H}$ with the energy $E$, as follows from Eq.~(\ref{TR_Ham}). Acting with ${\cal T}^2$ from the left,
\begin{equation}
{\cal T}^2 {\cal H}|E\rangle = E {\cal T}^2|E\rangle =  E|E_{T^2}\rangle,
\end{equation}
where $|E_{T^2}\rangle = - |E\rangle$ for the system of fermions. Thus,
\begin{equation}
- \langle E_T | E \rangle = \langle E_T | E_{T^2}\rangle =  \langle E_T | E \rangle,
\end{equation}
which implies that $|E\rangle$ and $|E_T\rangle \equiv {\cal T}| E\rangle$ are orthogonal and correspond to the same energy, proving degeneracy.
\end{proof}

\subsection{Furry's theorem}
\label{Furry}

Furry’s theorem is an important result in QED that stems from charge conjugation invariance. It forbids interactions involving an odd number of photons in fermion loops, significantly influencing the structure of higher-order QED processes. This theorem makes theoretical calculations easier and is crucial for explaining experimental findings related to particle decays and radiative corrections.

\begin{theorem}[Furry's theorem]
\textit{The sum of all Feynman diagrams containing an odd number of external photons attached to a single fermion loop is zero.}
\end{theorem}

\begin{proof}
Mathematically, the statement of the theorem is equivalent to the claim that the vacuum expectation value $\langle 0 |...| 0 \rangle$ of any odd number of electromagnetic currents vanishes, i.e. 
\begin{equation}\label{CorFun}
M_{2n+1}^{\mu_1 \mu_2 ... \mu_{2n+1}} = \langle 0 |T\left[j^{\mu_1} (x_1) j^{\mu_2} (x_2)...j^{\mu_{2n+1}} (x_{2n+1})  \right] | 0 \rangle = 0,
\end{equation}
where $j^\mu = \overline \psi \gamma^\mu \psi$ is the electromagnetic current. The theorem can be proven using the charge symmetry of QED. Recall that the QED vacuum is symmetric under charge conjugation, i.e. 
\begin{equation}
{\cal C} | 0 \rangle =  |0 \rangle, \ \mbox{and} \  \langle 0 | {\cal C}^\dagger =   \langle 0|.
\end{equation}
As was shown in the previous section, the vector current actually changes sign under the application of ${\cal C}$,
\begin{equation}
j^\mu \to {\cal C} j^\mu {\cal C}^\dagger  = - j^\mu.
\end{equation}
Inserting ${\cal C}^\dagger {\cal C}= 1$ into Eq.~(\ref{CorFun}) yields
\begin{eqnarray}\label{CorFun1}
M_{2n+1}^{\mu_1 \mu_2 ... \mu_{2n+1}} &=& \langle 0 |T\left[{\cal C}^\dagger {\cal C} j^{\mu_1} (x_1) {\cal C}^\dagger {\cal C} j^{\mu_2} (x_2)...{\cal C}^\dagger {\cal C}j^{\mu_{2n+1}} (x_{2n+1})  {\cal C}^\dagger {\cal C} \right] | 0 \rangle \nonumber \\
&=& \left(-1\right)^{2n+1} \langle 0 |T\left[j^{\mu_1} (x_1) j^{\mu_2} (x_2)...j^{\mu_{2n+1}} (x_{2n+1})  \right] | 0 \rangle = 0,
\end{eqnarray}
which shows that $M_{2n+1}^{\mu_1 \mu_2 ... \mu_{2n+1}}$ is equal to its negative because $2n+1$ is an odd number. This proves the conjecture: all diagrams with an odd number of external photons attached to a fermion loop vanish.
\end{proof}
\begin{corollary}
A single photon cannot be emitted from the vacuum or be absorbed by it.
\end{corollary}
%

\subsection{CP-violation in the Standard Model}
\label{Flavor}

It is experimentally established that CP represents a broken symmetry in the Standard Model. However, it is the {\it pattern} of CP symmetry breaking in the SM that makes its study interesting for New Physics searches. By examining the Lagrangian of the Standard Model, one can identify several potential areas where CP violation may occur \cite{Petrov:2021idw}. 

First, it is interesting to examine the terms of the SM Lagrangian that are {\it absent}. That is to say, if one writes all possible terms of dimension four or less built out of the SM fields, terms like
\begin{equation}\label{QCDtheta}
{\cal L} =
\frac{g^2  \theta}{32 \pi^2} G_{\mu\nu} \widetilde G^{\mu\nu},
\end{equation}
where $\widetilde G^{\mu\nu} = (1/2) \epsilon^{\mu\nu\alpha\beta} G^{\alpha\beta}$ is the dual field tensor, are conspiciously absent. 

One can argue that similar terms like $B \widetilde B$ for the abelian gauge fields can be written in terms of a full derivative and thus removed from the action \cite{Itzykson:1980rh}, while the weak SU(2) terms like $W \widetilde W$ do not have any observable consequences \cite{Anselm:1993uj}. Yet, the QCD term in Eq.~(\ref{QCDtheta}) can not be easily removed.

In fact, the presence of this term leads to an observable effect: a non-zero value of the electric dipole moment (EDM) of the neutron $d_n$. Experimentally, neutron's EDM is known to be very small, $\left|d_n\right| < 10^{-24}$ e$\cdot$cm. This implies that $\theta \ll 1$ is typically seen as evidence that CP is a good symmetry of strong interactions, making such terms unnecessary. While this "just so" explanation may be somewhat unsatisfactory, it offers an intriguing approach to modifying the minimal SM Lagrangian by incorporating additional dimension-four operators. 

Second, one could break CP-symmetry the same way we broke chiral symmetry, i.e., spontaneously. This could be achieved by acquiring complex phases in the broken phase. The vacuum expectation value of the Higgs field would then take the form
\begin{equation}
H=\frac{1}{\sqrt{2}}
\left( 
\begin{array}{c}
0\\
v e^{i \theta} 
\end{array} \right) .
\end{equation}
This mechanism could lead to observable CP-violating effects, but not within the minimal Standard Model. As can be seen from Eq.~(\ref{scalCPT}), the Higgs field transforms under a CP-symmetry transformation as 
\begin{equation}
\left[CP\right] H(t,{\mathbf r}) \left[CP\right]^\dagger = e^{i \alpha} H^\dagger(t, -{\mathbf r}),
\end{equation}
with $\alpha=\alpha_p + \alpha_c$. Thus, by choosing $\alpha=2\theta$, one can always make the SM vacuum invariant under CP transformation. Therefore, CP violation cannot occur spontaneously in the Standard Model. However, it can be realized in beyond-the-standard-model (BSM) frameworks featuring multiple Higgs doublets, representing yet another possible extension of the Standard Model.

Finally, a mechanism exists that encodes CP violation in the Standard Model. By considering the Yukawa terms of the SM Lagrangian, it indicates that this is the only place where CP can be broken in the SM Lagrangian. 

A rough argument is as follows: the Yukawa term in the Standard Model Lagrangian is  
\begin{equation}\label{SMYukawa}
{\cal L}_Y = Y_{ij} \overline \psi_{Ri} \chi_{Lj} \phi + Y_{ij}^\dagger \overline \chi_{Lj} \psi_{Ri} \phi^\dagger,
\end{equation}
where Eq.~(\ref{SMYukawa}) explicitly shows the Hermitian-conjugated part. From Table \ref{table2} and Eq.~(\ref{scalCPT}), the CP-conjugated Lagrangian is
\begin{equation}
\left[CP\right]  {\cal L}_Y  \left[CP\right]^\dagger = Y_{ij} \overline \chi_{Lj} \psi_{Ri} \phi^\dagger  + 
Y_{ij}^\dagger \overline \psi_{Ri}  \chi_{Lj} \phi.
\end{equation}
Comparing it to Eq.~(\ref{SMYukawa}), one can see that, unless the matrix of the Yukawa couplings is real, CP is broken. This is precisely what happens in the SM. The physical effect of the rotation from the gauge to mass basis can be seen in charged weak currents described by the term in the SM Lagrangian
\begin{equation}
{\cal L}_W = \frac{g_2}{\sqrt{2}} \overline u_{Li} \gamma^\mu 
\left[V_{uL} V_{dR}\right]_{ij} d_{Lj} W^+_\mu + h.c. = 
\frac{g_2}{\sqrt{2}} V_{ij} \ \overline u_{Li} \gamma^\mu d_{Lj} W^+_\mu + h.c.
\end{equation}
Here, $V_{ij}$ is the Cabbibo-Kobayashi-Maskawa (CKM) matrix written in terms of the Yukawa couplings in Eq.~(\ref{SMYukawa}). It has a generic form
\begin{equation}\label{CKMme}
V = 
  \left( {\begin{array}{ccc}
   V_{ud} & V_{us} & V_{ub} \\
   V_{cd} & V_{cs} & V_{cb} \\
   V_{td} & V_{ts} & V_{td}  \\
  \end{array} } \right).
\end{equation}
While the numerical values of the CKM matrix elements within the SM cannot be predicted, the number of parameters required for its complete parameterization can be established. To illustrate how this argument works, it may be helpful to consider a generic case of $N$ generations of the SM fermions, where the CKM matrix is a square $N \times N$ matrix.

First, a generic complex $N\times N$ matrix contains $2N^2$ real parameters. Since the CKM matrix is unitary, the unitarity relation $V V^\dagger = 1$ gives $N^2$ relations among the matrix parameters, i.e., sums of the products of matrix elements that are either $0$ or $1$. This results in $2 N^2-N^2=N^2$ angles or phases parameterizing the matrix. Further, since quark wave functions are only defined up to a phase, we can rotate the phases of up and down quarks to remove additional $2N-1$ parameters\footnote{There is one less than $2N$ rephasings needed because $V$ is always multiplied by a quark bilinear, which means that for at least one matrix element it is sufficient to rephase only one quark field. This is evident in the case of one up quark and one down quark: the CKM phase is eliminated by rotating the phase of {\it either} up or down quarks.}. This means that there remains only $N^2-(2N -1) = (N-1)^2$ parameters of a CKM matrix for $N$ generations of quarks.  

One can figure out how many of those parameters are angles. In the case of an $N$-dimensional space, one can perform $_NC_2=N(N-1)/2$ rotations in each plane, which means that the rest of the parameters are phases numbering $(N-1)^2 - N(N-1)/2=(N-1)(N-2)/2$. 

To summarize, for $N$ generations of quarks, there are 
\begin{eqnarray}
 \frac{N(N-1)}{2} && \ \ \mbox{angles} 
\nonumber \\
(N-1)(N-2)/2  && \ \ \mbox{phases}. 
\end{eqnarray}
This implies that for the case of $N=2$ generations, the CKM matrix has no phases, i.e., it is a real matrix. Now, for the $N=3$ generations of quarks, which is favored by Nature, there are three angles and one phase that completely parameterize the CKM matrix, in which case the CKM matrix alone can encode CP violation in the Standard Model. The ``standard'' parameterization of the CKM matrix can be written in terms of those angles and a phase,
\begin{eqnarray}\label{CKMStandard}
V &=& \begin{pmatrix}
1 & 0 &  0\\ 
0 & c_{23} & s_{23} \\ 
0 & -s_{23} & c_{23}  
\end{pmatrix}
\begin{pmatrix}
c_{13} & 0 &  s_{13} e^{-i\delta} \\ 
0 & 1 & 0 \\ 
-s_{13}e^{i\delta} & 0 & c_{213}  
\end{pmatrix}
\begin{pmatrix}
c_{12} & s_{12} &  0\\ 
-s_{12} & c_{12} & 0 \\ 
0 & 0 & 1  
\end{pmatrix}
\nonumber \\
&=&
\begin{pmatrix}
c_{12} c_{13} & s_{12} c_{13} &  s_{13} e^{-i\delta}\\ 
-s_{12} c_{23} - c_{12} s_{13} s_{23} e^{i\delta} & c_{12} c_{23} - s_{12} s_{13} s_{23} e^{i\delta} & c_{13} s_{23} \\ 
s_{12} s_{23} - c_{12} s_{13} c_{23} e^{i\delta} & -c_{12} s_{23} - s_{12} s_{13} c_{23} e^{i\delta} & c_{13} c_{23}  
\end{pmatrix}, \phantom{qwert}
\end{eqnarray}
where $c_{ij} = \cos\theta_{ij}$, $s_{ij} = \sin\theta_{ij}$, and $\delta$ is the CKM phase.

This is a rather remarkable result! It implies that all CP-violating effects in the Standard Model are related to a single phase of the CKM matrix! It also reinforces the fact that studies of CP violation are important tools for searches for New Physics, as any inconsistency of the observed CP-violating signals with the CKM picture would automatically lead to the discovery of physics beyond the Standard Model.

Although the numerical values of the CKM matrix elements cannot be predicted, they can still be measured experimentally. It is worth noting that the CKM matrix appears to have a very special form, with the diagonal matrix elements being nearly of order one and a gradual decrease in size away from the main diagonal. In fact, L. Wolfenstein proposed a very convenient way to parameterize the CKM matrix \cite{Wolfenstein:1983yz}, which resembles a perturbative expansion in parameter $\lambda \simeq |V_{us}| \sim 0.23$,
\begin{equation}\label{CKMWolfenstein}
V = 
  \left( {\begin{array}{ccc}
   1-\frac{\lambda^2}{2} & \lambda & A \lambda^3 (\rho - i \eta) \\
   -\lambda & 1-\frac{\lambda^2}{2} & A \lambda^2 \\
   A \lambda^3 (1-\rho - i \eta) & -A \lambda^2 & 1  \\
  \end{array} } \right) + {\cal O}(\lambda^4),
\end{equation}
where the other parameters are $A \simeq 0.81$, $\rho \simeq 0.14$, and $\eta \simeq 0.35$. They can be related to the standard parameterization of Eq.~(\ref{CKMStandard}) as $\lambda=s_{12}$, $A\lambda^2 = s_{23}$, and $A \lambda^3(\rho-i\eta) = s_{13} e^{-i\delta}$.

The current values of $\lambda$, $A$, $\rho$, and $\eta$ can be found in the Review of Particle Physics produced by the Particle Data Group \cite{ParticleDataGroup:2024cfk}. The parameterization of Eq.~(\ref{CKMWolfenstein}) is called the {\it Wolfenstein parameterization}\footnote{A variant of the parameterization in Eq.~(\ref{CKMWolfenstein}) is often used and is sometimes referred to as {\it Buras-Wolfenstein} parameterization. It can be obtained from Eq.~(\ref{CKMWolfenstein}) by substituting $\rho \to \bar \rho = \rho(1-\lambda^2/2)$ and $\eta \to \bar \eta=\eta(1-\lambda^2/2)$.}. It provides a very convenient way to estimate the sizes of different semileptonic and nonleptonic transitions based on their scaling with powers of $\lambda$. 

\subsection{Observing CP- and T-violation}
\label{Observables}
 
Experimental studies on the breaking of discrete symmetries and their combinations, particularly CP-violation, constitute a significant component of the experimental programs at the Large Hadron Collider (LHC) and flavor factories.  The results and methods are comprehensively detailed in various review papers and books \cite{Branco:1999fs,Bigi:2000yz,Petrov:2021idw}. The reader is directed to the extensive scientific literature on this subject. Nonetheless, a general discussion of these studies is warranted.

\subsubsection{Experimental studies of CP-violation}
\label{CPV_Exp}

Observables sensitive to CP violation are typically expressed in terms of asymmetries
\begin{equation}\label{CPV}
      A_{CP} = \frac{\Gamma(P \to f) - \Gamma(\bar{P} \to \bar{f})}{\Gamma(P \to f) + \Gamma(\bar{P} \to \bar{f})}
\end{equation}
formed from the partial decay rates of a state $P$ (which can be a meson or a baryon) to a final state $f$ and of its CP-conjugated counterpart, denoted by a bar. Depending on the initial state, the asymmetry in Eq.~(\ref{CPV}) could be a function of time. While the most compelling signals of CP-violation have been observed in nonleptonic transitions, strong interaction effects make it challenging to make precise predictions of such signals, so care must be taken in interpreting the experimental results. Useful experimental observables and sources of CP-violation allow it to classify them in three different categories \cite{Petrov:2021idw},
\begin{enumerate}
\item[(I)] 
For the neutral meson decays, CP violation can be observed in the meson-antimeson mixing matrix (``indirect'' CP violation). It is known that the introduction of $\Delta Q = 2$ transitions (such as $B^0-\bar B^0$ mixing), whether via the SM or beyond the SM—through one-loop or tree-level new physics amplitudes—leads to non-diagonal entries in the meson-antimeson mass matrix,
\begin{equation}
\label{MixingMatrix}
\left[M - i \frac{\Gamma}{2} \right]_{ij} = 
\left(
\begin{array}{cc}
A & p^2 \\
q^2 & A 
\end{array} 
\right)
\end{equation}
This type of CP violation is manifest when $R_m^2=\left|p/q\right|^2=(2 M_{12}-i \Gamma_{12})/(2 M_{12}^*-i 
\Gamma_{12}^*) \neq 1$.

\item[(II)]
For meson or baryon decays, CP violation can be seen in the $\Delta Q =1$ decay amplitudes (``direct'' CP violation). This type of CP violation occurs when the absolute value of the decay amplitude for a meson or baryon state to decay to a final state $f$ ($A_f$) is different from that of the corresponding CP-conjugated amplitude. This can happen if the decay amplitude can be broken into at least two parts associated with different weak and strong phases,
\begin{equation}\label{DirectAmpl}
A_f =
\left|A_1\right| e^{i \delta_1} e^{i \phi_1} +
\left|A_2\right| e^{i \delta_2} e^{i \phi_2},
\end{equation}
where $\phi_i$ represent weak phases ($\phi_i \to -\phi_i$ under CP-transormation), and $\delta_i$ represents strong phases ($\delta_i \to \delta_i$ under CP-transformation). This ensures that the CP-conjugated amplitude, $\overline A_{\overline f}$ would differ from $A_f$.

\item[(III)] CP violation in the interference of decays with and without mixing. This type of CP violation is possible for a subset of final states where neutral meson and antimeson can decay. 
\end{enumerate}
%

\subsubsection{Experimental studies of T-violation}
\label{T_Exp}

Experimental studies of T-violation serve as an important tool in the search for new physics. Since any local Lorentz-invariant quantum field theory conserves CPT,  T-violation consequently implies CP-violation. T-violating interactions are typically observed in T-odd asymmetries in triple-product correlations (TPCs). These correlations involve quantities of the form
\begin{equation}\label{TAsymm}
\mathcal{S} = {\mathbf m}_1 \cdot ({\mathbf m}_2 \times {\mathbf m}_3),
\end{equation}
where the vectors ${\mathbf m}_i$ are typically momenta or spin vectors of final-state particles in a decay process. Under time reversal, momenta, and spins change sign, so $\mathcal{S}$ is T-odd, as it contains an odd number of these quantities. It must be added that non-vanishing asymmetries of the type described in Eq.~(\ref{TAsymm}) could also be generated by final state interactions \cite{Zhitnitsky:1980he}, so care must be taken in extracting the {\it genuine} T-violating asymmetry. The procedure for doing so is typically as so:

First, the asymmetry is defined as
\begin{equation}
A_T = \frac{N(\mathcal{S} > 0) - N(\mathcal{S} < 0)}{N(\mathcal{S} > 0) + N(\mathcal{S} < 0)}.
\end{equation}
Second, a similar asymmetry is formed for the CP-conjugated process,
\begin{equation}
\bar{A}_T = \frac{N(\bar{\mathcal{S}} > 0) - N(\bar{\mathcal{S}} < 0)}{N(\bar{\mathcal{S}} > 0) + N(\bar{\mathcal{S}} < 0)}.
\end{equation}
Then, a genuine CP-violating observable is formed as
\begin{equation}
\mathcal{A}_{T\text{-odd}} = \frac{1}{2}(A_T - \bar{A}_T).
\end{equation}

Several examples are in order. For instance, in $\Lambda_b \to \Lambda \pi^+ \pi^-$ decay, a T-odd triple product can be constructed using the momenta of the final-state particles,
\begin{equation}
\mathcal{S} = {\mathbf p}_\Lambda \cdot ({\mathbf p}_{\pi^+} \times {\mathbf p}_{\pi^-}).
\end{equation}
This observable can be used to construct the asymmetry $A_T$ and compare it with its CP-conjugate process. Similarly, in $B \to V_1 V_2$ transitions (such as $B \to \phi K^*$), where both final-state mesons are vector mesons, a T-odd observable involves their polarization vectors,
\begin{equation}
\mathcal{S} = {\mathbf p}_B \cdot ({\mathbf \epsilon}_1 \times {\mathbf \epsilon}_2).
\end{equation}
These triple products appear in the angular distributions of the decay and are accessible through angular analysis. Finally, semileptonic decays such as $K^+ \to \pi^0 \mu^+ \nu_\mu$, a T-odd observable involves the transverse polarization of the muon,
\begin{equation}
\mathcal{S} = {\mathbf s}_\mu \cdot ({\mathbf p}_{\pi^0} \times {\mathbf p}_\mu).
\end{equation}
Such measurements are sensitive to CP-violating effects beyond the Standard Model since the SM contribution is highly suppressed. Similar constructions are also available for studies of parity violation in nuclear processes \cite{Bunakov:1982is}.

Triple-product correlations provide a clean and effective way to probe T-odd and CP-violating effects in various decay channels. Studying T-violating asymmetries complements other CP violation searches and is essential in exploring physics beyond the Standard Model.

\section{Conclusions}
\label{sec:conclusions}

Theoretical and experimental studies of C, P, and T symmetries, along with the patterns of their breaking, reveal fundamental aspects of particle physics. They elucidate various theoretical constructions and impose stringent constraints on the possible effects of physics beyond the Standard Model. The experimental observation of CP violation confirms the Standard Model and offers a glimpse into potential new physics. From kaons to B mesons, the pursuit of CP violation continues to illuminate the fundamental symmetries of nature.

\begin{ack}[Acknowledgments]%
 This work was supported in part by the US Department of Energy grant DE-SC0024357.
\end{ack}


\bibliographystyle{Numbered-Style} 
\bibliography{CPV_reference}

\end{document}